\documentstyle[twocolumn,floats,aps]{revtex}

\input epsf         
\epsfverbosetrue    
\def\postscript#1{\begin{center}\leavevmode
\hbox{\epsfxsize=0.95\columnwidth\epsfbox{#1}}\end{center}}

\begin{document}

\twocolumn[\hsize\textwidth\columnwidth\hsize\csname@twocolumnfalse%
\endcsname


\title{Commensurate and Incommensurate Spin Fluctuations in
YBa$_2$Cu$_3$O$_{6+y}$}

\author{Khee-Kyun Voo and W. C. Wu}
\address{Department of Physics, National Taiwan Normal University,
Taipei 11650, Taiwan}

\date{\today}

\maketitle

\begin{abstract}

We present an interpretation of the recent neutron data on the
commensurate and incommensurate spin fluctuations found in
YBa$_2$Cu$_3$O$_{6+y}$ based on a special configuration of the
electronic dispersion and intervention from the $d_{x^2-y^2}$-wave
superconducting phase. The observed switch over between the
commensurate and incommensurate fluctuation spectra at the change
of frequency or temperature is naturally accounted within this
scenario.

\end{abstract}

]

Recent inelastic neutron scattering (INS) experiments on
YBa$_2$Cu$_3$O$_{6+y}$ (YBCO) compounds show a lot of new boggling
data in spin dynamics. Since magnetic fluctuations in the
high-$T_c$ cuprates are long thought to be intimately related to
their mechanism of superconductivity, it has attracted very much
attention.


We briefly review the INS experiments which measure spin
susceptibility Im$\chi_s$ on YBCO. The INS in low-temperature
superconducting (SC) phase optimal-doped YBCO shows a peak at
frequency $\omega_0(T$$=$$0)=$ 41meV and momentum ${\bf
Q_\pi}=(\pi,\pi)$ \cite{Rossat91}. The frequency width of the peak
is within 10meV and momentum width is within $0.1\pi\sim0.5\pi$
\cite{Bourges96,Fong96}. In underdoped YBCO, the resonance
frequency is slightly softened, and some damping is developed both
in the frequency and momentum space \cite{Fong97}. When the
temperature is raised, the resonance frequency $\omega_0(T)$ is
softened only very little and intensity slightly suppressed before
it abruptly disappears at the transition to normal state
\cite{Bourges96}. Recently it was found that the incommensurate
fluctuation, which was found previously only in
La$_{2-x}$Sr$_x$CuO$_4$ (LSCO) compounds, also exist in YBCO
\cite{Dai98,Arai99}. The spin fluctuation in low-temperature YBCO
at frequencies either lower or higher than $\omega_0(0)$ is
incommensurate in nature, characterized by four peaks at
$(\pi,\pi\pm\delta)$ and $(\pi\pm\delta,\pi)$ as in the case of
LSCO. At frequencies just below $\omega_0(0)$ the deviation of the
incommensurate peak from ${\bf Q_\pi}$ is also found to be
slightly less than those at lower frequencies, where in contrast
the deviation grows with frequency at $\omega>\omega_0(0)$
\cite{Arai99}. There is also evidence that at frequency
substantially below $\omega_0(0)$, raising the temperature from
below to high above $T_c$ brings the fluctuation from
incommensurate to commensurate \cite{Dai98}. Contradictory, in a
similar underdoped YBCO, a weak incommensurate structure is found
recently by Arai $et$ $al$ \cite{Arai99} at a similar high
temperature regime.

On the other hand, a similar commensurate peak was also found very
recently in Bi$_2$Sr$_2$CaCu$_2$O$_8$ (BSCCO) \cite{Fong99}.
Though less definite, an incommensurate fluctuation also occurs in
these compounds at lower frequencies \cite{Mook98}

Since in earlier times commensurate fluctuation was specific in
YBCO while incommensurate in LSCO, the fluctuations were believed
to be mutually unrelated and hence theoretical approaches to them
were independent. The recent found interrelation thus sets a new
constraint on the candidate theory of these systems.


The incommensurate peaks are believed as, due to Fermi surface
nesting in one form or another \cite{Lu92,Brinckmann99}. Whereas
for the commensurate peak, various possibilities were proposed.
Some focused on its peak feature in frequency space
\cite{Blumberg95,Bulut96,Yin97,Abrikosov98} while others gave more
comprehensive treatments including the overall behavior of
Im$\chi_s$ in momentum space
\cite{Brinckmann99,Demler95,Liu95,Mazin95,Morr98,Kao99}. Because
of the recent discovery of incommensurate fluctuation in YBCO,
unified treatments of the fluctuations were also put forward
\cite{Brinckmann99,Kao99}. A common feature of these theories is
that most of them incorporate some interaction effect into the
susceptibility via Random-Phase-Approximation (RPA) in some form
to result in a collective mode at ${\bf Q}_\pi$
\cite{Brinckmann99,Blumberg95,Demler95,Liu95,Mazin95,Kao99}. In
this Letter we address the mentioned observations to a property of
the band structure in conjunction with the intervention from the
$d_{x^2-y^2}$-wave SC phase. Thus in contrast to the collective
mode via RPA scenarios, our view of the 41meV peak is based on the
bare susceptibility. Furthermore we also include the discussion of
the switch-over between commensurate and incommensurate
fluctuations as driven by temperature, which so far absent from
the literature.


Our approach was initiated by recent angle resolved photoemission
spectroscopy (ARPES) measurement on YBCO and BSCCO. The
measurement shows the existence of an extended van Hove
singularity (VHS) at the ${\overline M}$ points [$(0,\pm\pi)$ and
$(\pm\pi,0)$] of the Brillouin zone situated within 20meV below
the Fermi level \cite{Dessau93,Gofron94}. The authors relates this
to the occurrence of higher $T_c$ of these compounds as enhanced
by the VHS \cite{Gofron94}, compared to NdCeCuO compounds where
the VHS is found hundreds of meV below the Fermi level
\cite{King93}. On the other hand, both YBCO and BSCCO were found
to exhibit commensurate fluctuation, and incommensurate
fluctuation at lower frequencies. We believe that these common
features are not accidental. Therefore in this paper we start out
from an electronic dispersion with Fermi level near the VHS at
${\overline M}$ to describe the INS experiment on YBCO. Previously
there were also interpretations of the commensurate peak based on
the VHS effect. But the authors considered the RPA-type
susceptibility in bilayer systems \cite{Blumberg95,Mazin95} and a
$s^{\pm}$-wave gap \cite{Mazin95}. A mono-layer VHS scheme closer
to ours also exist but the discussion on the bare susceptibility
was limited to zero temperature and specific to commensurate peaks
\cite{Abrikosov98}.

Our description of the spin fluctuations in YBCO is in good
agreement with experiments. The absence of commensurate magnetic
response in LSCO is probably due to the remoteness of Fermi level
from the VHS at ${\overline M}$ as it has a lower $T_c$.


We start with the single-band bare spin susceptibility

\begin{eqnarray*}
&&\chi_s^0({\bf q},\omega)=-{1\over 4}\sum_{\bf k}\left[\left[
1-{\varepsilon_{\bf k}\varepsilon_{\bf k+q}+\Delta_{\bf
k}\Delta_{\bf k+q} \over E_{\bf k}E_{\bf k+q}}\right]\right. \nonumber\\
&&\times \left[\left.{1-f(E_{\bf k})-f(E_{\bf k+q})\over
\omega-E_{\bf k}-E_{\bf k+q}+i\delta}-
{1-f(E_{\bf k})-f(E_{\bf k+q})\over \omega+E_{\bf k}+E_{\bf k+q}+i\delta}
\right]\right. \nonumber\\
&&-\left[\left.  1+{\varepsilon_{\bf k}\varepsilon_{\bf k+q}+
\Delta_{\bf k}\Delta_{\bf k+q} \over E_{\bf k}E_{\bf k+q}}\right]\right.
\nonumber\\
&&\times \left[\left. {f(E_{\bf k})-f(E_{\bf k+q})\over
\omega-E_{\bf k}+E_{\bf k+q}+i\delta}-
{f(E_{\bf k})-f(E_{\bf k+q})\over \omega+E_{\bf k}-E_{\bf k+q}+i\delta}
\right]\right],
\label{eq:chis0}
\end{eqnarray*}
where $f(E_{\bf k})$ is the Fermi function and $E_{\bf
k}=(\varepsilon_{\bf k}^2+\Delta_{\bf k}^2)^{1\over 2}$ is the
quasiparticle excitation spectrum with $\varepsilon_{\bf k}$ and
$\Delta_{\bf k}$ the electronic dispersion and superconducting gap
respectively. We use a two-dimensional tight-binding electronic
dispersion

\[
\varepsilon_{k}=-2t(\cos k_{x}+\cos k_{y}) -4t^{'}\cos k_{x}\cos
k_{y}-\mu,
\label{eq:tb}
\]
where $t$ and $t^\prime$ are the nearest-neighbor (NN) and
next-nearest-neighbor (NNN) hopping respectively. We have fixed
$t'=-0.178t$ throughout this paper and $\mu$ is the chemical
potential that depends on doping. With YBCO in mind, the gap is
taken as $\Delta_{\bf k}=\Delta(T)(\cos k_{x}-\cos k_{y})/2$. The
maximum gap at zero temperature is taken as $\Delta(0)=0.3t$ and
this point will be further elaborated. The numerical integration
width is taken as $\delta=0.004t$ and the slicing is
$2000\times2000$.

The chosen electronic dispersion is meant to phenomenologically
describe the empirical relation between the Fermi level and the
VHS. The chosen NNN hopping $t^{'}=-0.178t$ has a Fermi surface
crossing ${\overline M}$ at doping level $x=0.15$ [Fig.1(b)],
which is here nominally taken as the optimal doped case. This is
to simulate a most intense effect of the VHS at an optimal doped
compound as to result at a highest $T_c$. We choose $x=0.10$
throughout this paper unless otherwise stated, as to show that the
commensurate peak exist even when the Fermi surface is $not$
exactly situated at the VHS but sufficiently close to it. And we
believe that the only crucial point is the small difference in
energy between the VHS and Fermi level but $not$ their distance in
momentum space as seen in the ARPES experiment. Whatever the
origin of the VHS is also out of our concern. Furthermore we adopt
a rigid band approximation. As far as the chosen dispersion has a
Fermi surface out-bowed from the origin of the Brillouin zone as
seen in the ARPES experiment, the main conclusion of our
discussion is believed to valid.


In Fig.1(a), the frequency dependence of zero-temperature
Im$\chi_s^0({\bf Q_\pi},\omega)$ at several doping levels ranging
from underdoped to overdoped are shown to exhibit a peak near
$2\Delta(0)$. Im$\chi_s^0({\bf Q_\pi},\omega)$ is gapped out at
low frequencies since ${\bf Q_\pi}$ spans regions of maximal gap.
It shows a high intensity at a small window around $2\Delta(0)$
due to the enhancement by the coherence factor on transitions
between opposite gap phase regions, and is most intense when the
frequency is just enough to open up the transition [which is near
$2\Delta(0)$]. Furthermore, these regions are situated near
${\overline M}$. Thus the peak arises due to a co-enhancement from
the VHS and coherence factor. Note that since the gap is written
in term of the NN-hopping, which is generally different from
system to system, the absolute resonance frequencies and
linewidths of different doped cases could not be directly
compared. But if we phenomenologically input the experimental fact
that $\omega_0(0)$ is only moderately decreased at underdoping,
the peak could be argued as always be broadened at underdoping. We
give the argument as follows. Denoting $\Gamma$ as the half-height
width of the peak. At doping $x=0.05$, with choices of
$\Delta(0)/t$ within 0.2$\sim$0.5, the ratio $\Gamma/\omega_0(0)$
runs within 0.14$\sim$0.07. While at $x=0.15$, the same choice of
$\Delta(0)/t$ give $\Gamma/\omega_0$ within 0.05$\sim$0.04 (in
both cases a smaller $\Delta(0)/t$ gives a larger
$\Gamma/\omega_0(0)$). It is seen that if the decrement of
$\omega_0$ is within 30 percent, the broadening in an absolute
energy scale always exist regardless of the choice of
$\Delta(0)/t$ in individual systems within a broad range.

\begin{figure}[t]
\vspace{-0.3cm}
\postscript{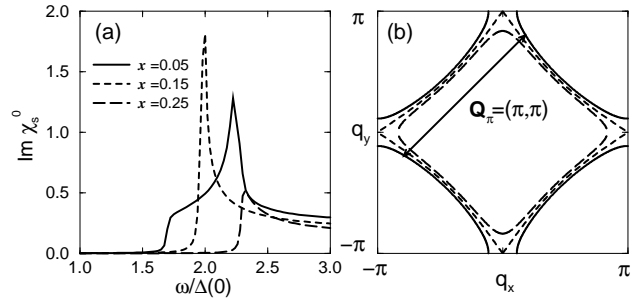}
\vspace{-0.3cm}
\caption{(a) Zero-temperature Im${\chi_s^0}({\bf Q_\pi},\omega)$
at doping $x=0.05, 0.15$, and $0.25$. When making comparison of
the resonance frequencies in an absolute energy scale, the
empirical fact of $\Delta(0)$ in underdoped YBCO could be smaller
should be taken into account. The broadening of the peak upon
underdoping is discussed in the text. (b) Fermi surfaces
correspond to the doping levels. The ${\bf Q_\pi}$ transition
corresponding to the leading edge of the peak at $x=0.05$ is
shown.} \label{Fig1}
\end{figure}

The broadening of the peak in frequency space is due to a mismatch
of both enhancement effects. As the compound is underdoped, the
${\bf Q_\pi}$ transition on the Fermi surface shifts away from the
full-gap regions [see Fig.1(b)] and leads to a down shift of the
leading edge of the peak. On the other hand the VHS remains to
affect only when the frequency is sufficient to open up the
transition to the ${\overline M}$ points. In the overdoped case,
Im$\chi_s^0({\bf Q_\pi},\omega)$ is suppressed because the Fermi
surface has shrunk to a size that ${\bf Q_\pi}$ is too long to
connect the Fermi surfaces, which all transitions are bound on
[Fig.1(b)].

Im$\chi_s^0({\bf q},\omega_0)$ in momentum space is indeed a peak
at ${\bf Q_\pi}$ with a reasonable width [Fig.2(c)]. In our
calculation the broadening effect of underdoping is more obvious
in frequency space than in momentum space. Another important point
to note is that apart from the nearness of the Fermi level to the
VHS, the existence of the peak also demand the existence of a
$d_{x^2-y^2}$-wave gap, but not a mono-layer $s$-wave gap.


\begin{figure}[t]
\begin{center}
\vspace{-1.0cm}
\leavevmode\epsfxsize=3.6in \epsfbox{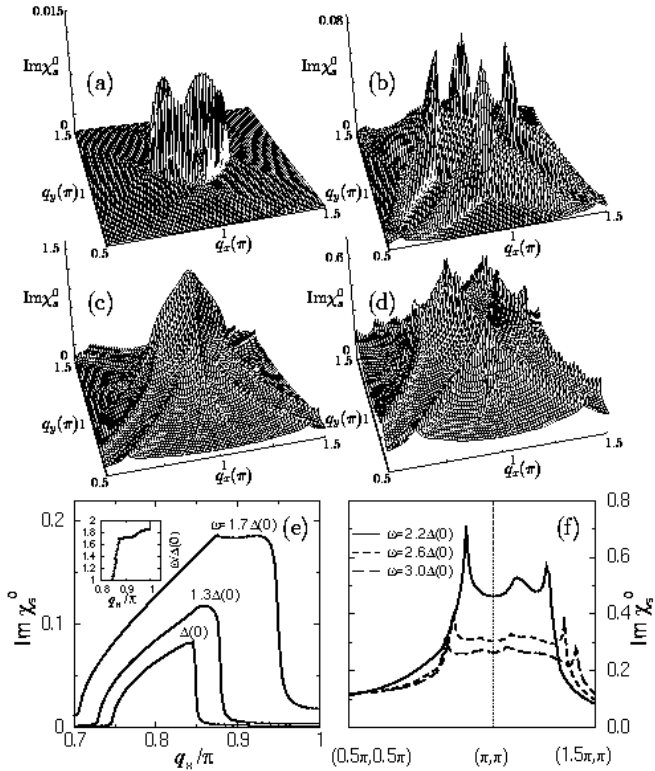}
\vspace{-0.4cm}
\end{center}
\caption{ The frequency evolution of Im${\chi_s^0}$ in momentum
space at $T=0$ and $x=0.10$. (a) $\omega=0.3\Delta(0)$; (b)
$\omega=1.0\Delta(0)$; (c) $\omega=2.025\Delta(0)$ $[i.e.$ $
\omega_0(0)]$; (d) $\omega=2.2\Delta(0)$. (e) Scans along ${\bf
q}=(0.7\pi,\pi)$$\rightarrow$$(\pi,\pi)$ at frequencies
$\omega/\Delta(0)=1.0, 1.3$, and $1.7$. The inset shows the
$\omega$-dependence of the momentum location for the tips of the
peaks. (f) Scans along ${\bf q}=
(0.5\pi,0.5\pi)$$\rightarrow$$(\pi,\pi)$$\rightarrow$$(1.5\pi,\pi)$
at $\omega/\Delta(0)=2.2, 2.6$, and $3.0$.} \label{Fig2}
\end{figure}

Fig.2 shows the zero-temperature Im$\chi_s^0$. It consist of
node-to-node incommensurate peaks at quasielastic frequencies,
Fermi-surface-nesting incommensurate peaks at intermediate
frequencies, commensurate peak at frequency $\omega_0(0)$ near
$2\Delta(0)$, and incommensurate peaks at frequencies over
$\omega_0(0)$ [Fig.2(a)-(d)]. The origin of the node-to-node and
nesting incommensurate peaks are well accounted in the literature
\cite{Lu92,Brinckmann99}. In our scenario such sequential
emergence of the peaks naturally occurs in contrast to the RPA
scenarios where the resonance frequency of the commensurate peak
could be sensitive to the choice of the interaction strength.

There is a decrease of the incommensurateness of the nesting
incommensurate peaks at frequencies just below $\omega_0(0)$
[Fig.2(e)]. This is due to a dynamical nesting effect. Since the
Fermi surface is out-bowed, the opened up energy contour $E_{\bf
k}=\omega/2$ which determines the possible transitions
\cite{Brinckmann99} has a smaller (larger) curvature above (below)
the Fermi surface. Thus local nesting is better between contour
above the Fermi surface and this determines the location of the
tip of the peak. As $\omega$ increases, the contour surges away
from the Fermi surface and that pushes the nesting peaks towards
${\bf Q_\pi}$.

At frequencies over $\omega_0(0)$, due to the filling up of the
valley at ${\bf Q_\pi}$ a hardly discernible incommensurate
feature is seen [Figs.2(d) and 2(f)]. It is characterized by
incommensurate peaks displaced from ${\bf Q_\pi}$ along both the
zone diagonals and edges. The incommensurateness grows with
frequency in this regime [Fig.2(f)] and that agrees qualitatively
with experiment \cite{Arai99}. Further comment will be made on
this point at the end of this Letter.


In our following discussion of the temperature evolution, we take
a convenient empirical relation between the gap and temperature
$\Delta(T)/\Delta(0)=[1-(T/T_c)^5]^{1/2}$ [see inset in Fig.3(a)]
and a typical ratio $2\Delta(0)/T_c=8$.

\begin{figure}[t]
\vspace{-0.5cm}
\postscript{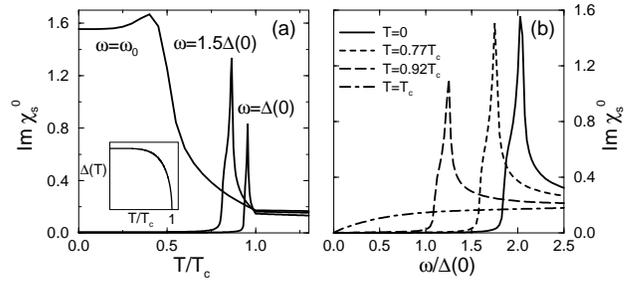}
\vspace{-0.3cm}
\caption{(a) Temperature dependence of Im${\chi_s^0}({\bf
Q_\pi},\omega)$ at $\omega=2.025\Delta(0)$ $[i.e.$ $\omega_0(0)],$
$1.5\Delta(0),$ and $1.0\Delta(0)$. (b) $\omega$-dependence of
Im${\chi_s^0}$ at ${\bf Q_\pi}$ and $T=0, 0.77T_c$, $0.92T_c$, and
$T_c$. The doping is fixed at $x=0.10$ and the temperature
dependence of gap $\Delta(T)$ is shown in the inset of (a).}
\label{Fig3}
\end{figure}

In Fig.3(a), Im$\chi_s^0$ at $({\bf Q_\pi},\omega_0(0))$ is seen
to be suppressed by temperature as in experiments
\cite{Bourges96}. At temperature below $T_c$, the main cause of
the suppression is the shift away of the mode frequency
$\omega_0(T)$ from $\omega_0(0)$, the diminishing of the peak
intensity is actually less rapid [Fig.3(b)]. For temperatures
$T<0.8T_c$, the mode frequency of the commensurate peak is
softened a little from $\omega_0(0)$ before abruptly softened to
zero and disappears at $T_c$. The softening is merely a
manifestation of the diminishing gap and owing to $T_c$ which is a
small energy scale compared to the bandwidth, the peak feature is
maintained well at $T<T_c$.


While the temperature evolution of the commensurate peak at
$\omega=\omega_0(0)$ has been frequently discussed
\cite{Bulut96,Yin97,Liu95}, the temperature evolution of the
incommensurate peaks at $\omega<\omega_0(0)$ has been overlook.
Nevertheless recent experiments show that this is not at all
trivial \cite{Dai98,Arai99}. At low frequency and temperature,
Im$\chi_s^0$ at ${\bf Q_\pi}$ is gapped out and leads to a clear
appearance of the incommensurate peaks. As the temperature is
raised, the particular frequency could equal to the temperature
dependent resonance frequency and the fluctuation is then
converted to commensurate at a $\omega<\omega_0(0)$ and $T\alt
T_c$ [compare Fig.2(b) and Fig.4(a)]. Plotting Im$\chi_s^0$ at the
$\omega$ and ${\bf Q_\pi}$ against the temperature shows the
emergence of the peak at a temperature just below $T_c$
[Fig.3(a)]. As the temperature is raised further into the normal
state, the fluctuation reverts to a very weak incommensurate
feature \cite{Voo99}. Experimentally it may well escape from
detection and seen as a broad commensurate structure. At further
underdoping, say $x=0.05$, even a well distinguished commensurate
feature may appear in the normal state [Fig.4(b)] and it is
suppressed as the temperature increases. Note that the intensity
of this high temperature commensurate structure is of the same
order of magnitude as its low temperature incommensurate
counterpart as seen in the experiment \cite{Dai98}. In our case,
the valley at ${\bf Q_\pi}$ appears and leads to a pronounced
incommensurate structure in the normal state only in the overdoped
case \cite{Voo99}. This may resolve the apparent contradiction
between the observation of a broad commensurate structure
\cite{Dai98}, and the observation of a weak incommensurate
structure \cite{Arai99} at similar temperature $(>T_c)$ and
frequency $(<\omega_0)$ regimes in an underdoped YBCO. In our
calculation these weak incommensurate structures are deviated from
${\bf Q_\pi}$ along the zone edges.


\begin{figure}[t]
\begin{center}
\vspace{-0.5cm}
\leavevmode\epsfxsize=3.7in \epsfbox{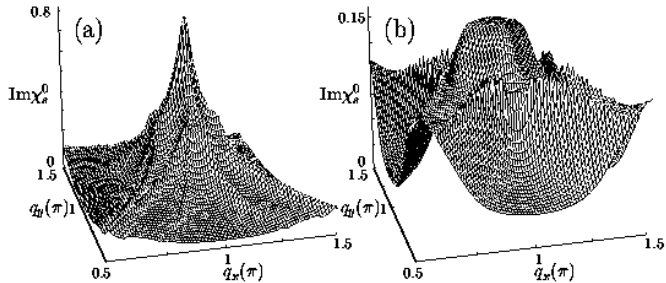}
\vspace{-0.2cm}
\end{center}
\caption{Finite-temperature Im${\chi_s^0}$ in momentum space at an
intermediate frequency $\omega=1.0\Delta(0)$. $\Delta(0)=0.3t$ and
$2\Delta(0)/T_c=8$ throughout. (a) $T=0.955T_c$ and $x=0.10$;  (b)
$T=T_c$ and $x=0.05$. } \label{Fig4}
\end{figure}

We make a last comment on our result. If a Hubbard repulsion of
$1.0t$ is incorporated via a RPA spin susceptibility, structure at
the vicinity of ${\bf Q_\pi}$ will be more prominent than
structures away from ${\bf Q_\pi}$. Furthermore, at zero
temperature a very sharp and intense resonance at ${\bf Q_\pi}$
occurs at $\omega_0=1.8\Delta(0)$. The incommensurate structure at
$\omega>\omega_0$ is prominently away from ${\bf Q_\pi}$ along the
Brillouin zone edges. If a smaller repulsion $0.5t$ is taken,
every structure resembles those of the bare susceptibility. Since
we have no reason to believe that the repulsion should be finely
$1.0t$, we believe that such a tuning is a less probable
description of reality.


In conclusion, we have provided an unified explanation of the INS
experiments on YBCO based on the closeness of the Fermi level to
the VHS at ${\overline M}$ points and coherence effect from the
$d_{x^2-y^2}$-wave SC phase. The results describe naturally the
frequency and temperature dependence of the commensurate peak, the
switch over between the incommensurate and commensurate spectra at
the change of frequency or temperature, and the decrease
[increase] of the incommensurateness at frequency
$\omega\alt2\Delta(0)$ [$\omega>2\Delta(0)$]. Furthermore, the
broadening in frequency space of the commensurate peak at
underdoping is also accounted semi-phenomenologically.


We thank JH Kok and YT Huang for helpful conversations and
acknowledge the support from NSC of Taiwan under grant
No.89-2112-M-003-009.

\end{document}